\documentstyle[pra,aps,twocolumn,epsf]{revtex}
\renewcommand{\vec}[1]{\mbox{\boldmath$#1$}}

\begin{document}
\draft
\twocolumn[\hsize\textwidth\columnwidth\hsize\csname @twocolumnfalse\endcsname

\title{
Theoretical Study of Pressure Effect on TDAE-C$_{60}$
}

\author{
Tohru Kawamoto,
Madoka Tokumoto, 
Hirokazu Sakamoto$^1$, and 
Kenji Mizoguchi$^1$
}

\address{
Electrotechnical Laboratory, 1-1-4 Umezono, Tsukuba 305-8568, Japan. \\
$^1$ Department of Physics, Tokyo Metropolitan University,
 Minamiosawa, Hachioji, Tokyo 192-0397.
}
\date{\today}
\maketitle

%
%

\begin{abstract}
We have theoretically studied pressure effects on 
molecular ferromagnet C$_{60}$ complexes 
with tetrakis (dimethylamino) ethylene (TDAE),
particularly the pressure-induced depression of the Curie temperature. 
The observed behavior is well simulated with our model 
based on a charge transfer induced intramolecular Jahn-Teller 
distortion and an intermolecular cooperative Jahn-Teller interaction. 
We emphasize that the theoretical simulation is carried out with 
reasonable parameters  known for C$_{60}^-$ complexes. 
It is concluded that the enhancement of crystal field at C$_{60}$ site 
due to increasing pressure 
causes the depression of Curie temperature. 
\end{abstract}
\vskip2pc]

In 1991, Allemand {\it et al. } reported
the ferromagnetic behavior of 
tetrakis (dimethylamino) ethylene (TDAE) -C$_{60}$ 
with $T_{\rm C}$=16 K\cite{Allemand91Science}. 
It has been attracting many scientists due to its having the 
highest Curie temperature 
among pure organic molecular ferromagnets and its unusual magnetism. 
Allemand {\it et al.} proposed `soft ferromagnetism' 
with no hysteresis in the M-H curve\cite{Allemand91Science}.
Tanaka {\it et al.}\cite{Tanaka93PRB} and Blinc {\it et al.}\cite{Blinc94SSC} 
have observed behaviors like a `superparamagnetism' of spin clusters 
consisting of hundreds of spins in the magnetization and 
the proton nuclear magnetic resonance (NMR) measurements 
below $T_{\rm C}$, respectively.
Venturini {\it et al.} suggested `spin glass model' from analysis 
of the electron spin resonance (ESR) lineshape\cite{Venturini92IJMPB}.
In 1997, we proposed a model 
for clarifying the properties of this system\cite{Kawamoto97SSC}. 
This model is introduced in order to  clarify
the origin of intermolecular ferromagnetic coupling between C$_{60}$'s, 
and it is based on the orbital ordering of unpaired electrons 
on C$_{60}$'s 
due to the adjacent alignment of the Jahn-Teller distorted  C$_{60}$'s. 
The spin-glass-like behavior and superparamagnetism of the spin clusters 
may also be explained qualitatively in this framework \cite{Kawamoto97SynthM}. 

\begin{figure}
\begin{center}
  \leavevmode
  \epsfxsize=80mm
  \epsfbox{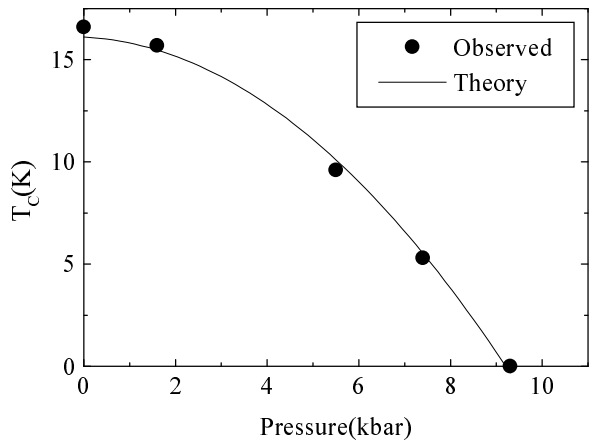}
\end{center}
\caption{
The pressure dependence of Curie temperature in TDAE-C$_{60}$. 
Closed circles represent observed value. 
Solid line indicates the simulation result based on the orbital ordering model. 
}
\label{fig:p_tc}
\end{figure}

Very recently, Mizoguchi {\it et al.} observed the pressure dependence 
of the Curie temperature ($T_{\rm C}$) in TDAE-C$_{60}$, as 
shown in Fig.~\ref{fig:p_tc}\cite{Mizoguchi00PRL}. 
It appears that $T_{\rm C}$ is parabolically depressed upon the application of 
pressure. 
The purpose of this paper is to provide 
a quantitative understanding of this behavior 
in the framework of the model mentioned above. 
It should be noted that the observed value is simulated quantitatively with 
the reasonable parameters known for the C$_{60}$ molecule. 

\begin{figure}
\begin{center}
  \leavevmode
  \epsfxsize=80mm
  \epsfbox{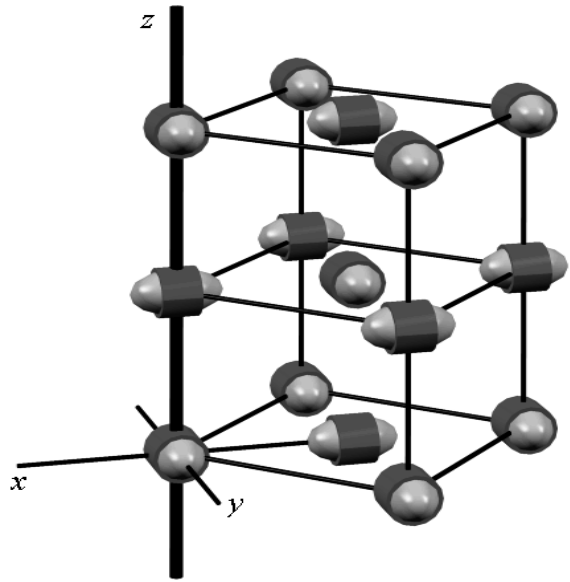}
\end{center}
\caption{
An example of possible JT--distorted crystal structures 
which are likely to cause a three-dimensional ferromagnetic ordering. 
Elongated spheres indicate JT--distorted 
C$_{60}$'s and thick lines express one-dimensional chains. 
Note that distortion is exaggerated. 
In this structure the elongated axes of C$_{60}$'s are perpendicular to 
each other for the interchain nearest neighbor C$_{60}$'s as well as for the 
intrachain nearest neighbor C$_{60}$'s. 
The gray belts around C$_{60}$'s schematically represent 
the distribution of a unpaired electron.
}
\label{fig:c60_jt}
\end{figure}

First, our orbital ordering model is briefly introduced. 
With respect to 
this model, the magnetic interaction between local spins on C$_{60}$'s
are discussed. 
In TDAE-C$_{60}$, both TDAE and C$_{60}$ are regarded to have 
an unpaired electron since one electron transfers from TDAE to C$_{60}$. 
However, it has not yet been clarified 
whether or not bare spin moments exist on TDAE molecules. 
On the other hand, 
unpaired electrons on C$_{60}$ molecules are considered to
play a crucial role in bulk ferromagnetism. 
Figure \ref{fig:c60_jt} shows an example of a molecular arrangement 
which is likely to cause three-dimensional ferromagnetic order. 

The lowest unoccupied orbitals (LUMO) of C$_{60}$ in I$_{\rm h}$ 
are triply degenerated with $t_{\rm 1u}$ symmetry. 
When degenerated orbitals are partially occupied, 
the molecule is distorted in order to stabilize one orbital, 
which is known as the Jahn-Teller (JT) effect. 
We assume the D$_{2h}$ structure for the C$_{60}^{-}$ anion in this study. 
We performed a geometrical optimization of single a C$_{60}^{-}$ anion 
in a previous study. 
The optimized structure resembles a rugby ball whose elongated axis lies 
along one of the three symmetry axes in the D$_{2h}$ structure. 
The atomic displacement from the ideal icosahedral 
is small, {\it i.e.}, 0.01~\AA~ at most.
The three t$_{1u}$ states, LUMO$_x$, LUMO$_y$, and LUMO$_z$, 
which are degenerate LUMO's of C$_{60}$ 
with I$_h$ symmetry before the charge-transfer, 
each split into three states.
If the elongated axis is the $x$-axis, 
LUMO$_x$ has the lowest energy, and 
the other two orbitals, LUMO$_y$ and LUMO$_z$, with higher energies 
are almost degenerate. 
It should be noted that the charge density is not spherical but 
takes large values along a  belt surrounding the elongated axis 
(see Fig.~\ref{fig:c60_jt}). 

Magnetic interactions between distorted molecules depend on 
their alignment. 
We concluded that the ferromagnetism is realized in 
the structure illustrated in Fig.~\ref{fig:c60_jt} 
due to periodic molecular distortion\cite{Kawamoto97SSC}. 
In a C$_{60}^{-}$ array along the $c$-axis in this structure, 
the elongated axes of two nearest C$_{60}^{-}$'s 
are perpendicular to each other. 
In this case, the orbitals of unpaired electrons align in an alternating 
manner, 
for example, LUMO$_{\rm x}$, LUMO$_{\rm y}$, LUMO$_{\rm x}$,...
Such a system is called an orbital ordering system 
and an interaction acting to favor an alternating alignment is called a
cooperative JT interaction. 
It is known that the ferromagnetic coupling between neighbors is preferred 
in such a orbital ordering system. 
It should be noted that the 
intermolecular transfer occurs only between the same kind of orbitals 
if the crystal field at the C$_{60}^{-}$'s has perfectly orthorhombic symmetry. 
It is a key point in favoring the intermolecular ferromagnetic interaction. 

It is expected that the magnetic interactions 
between C$_{60}$'s lying along the $c$-axis are the strongest 
among the intermolecular magnetic interactions, 
because their distance is shorter than that in the ab-plane by 0.3~\AA. 
Electrical transport along the $c$-axis is also observed about 10 times larger 
than that along the $a$-axis\cite{Omerzu97SynthM}. 
In this paper, 
we assume that the system is quasi-one-dimensional along the $c$-axis.
The extended Hubbard Hamiltonian for the one-dimensional chain 
with LUMO$_{\rm x}$'s and  LUMO$_{\rm y}$'s as the basis functions 
is shown as follows:
\begin{eqnarray}
H  & = & \frac{\Delta}{2} \sum_{i\mu\sigma} (-1)^i n_{ix \sigma }
                               +(-1)^{i+1} n_{iy \sigma } \label{eq:Hubbard} \\
           &+&t_{g}\sum_{i\sigma} 
               (c_{i x \sigma } ^{\dagger} c_{i+1 y \sigma } 
               +c_{i y \sigma } ^{\dagger} c_{i+1 x \sigma } 
                                                       + h.c.) \nonumber \\
           &+&t_{\ell}\sum_{i\sigma} 
               (c_{i x \sigma } ^{\dagger} c_{i+1 x \sigma } 
               +c_{i y \sigma } ^{\dagger} c_{i+1 y \sigma } 
                                                       + h.c.)  \nonumber \\
    & + & \sum_{i\mu} U n_{i\mu\uparrow}n_{i\mu\downarrow}  \nonumber \\
    & + & \sum_{i\mu\neq\nu\sigma\sigma^{\prime}}
            (U-J\delta_{\sigma\sigma^{\prime}})n_{i\mu\sigma}
                                      n_{i\nu\sigma^{\prime}}  \nonumber \\
    & + & \sum_{i\mu\nu\sigma}Jc_{i\mu \sigma } ^{\dagger} 
                 c_{i\nu\bar{\sigma}} ^{\dagger}
                 c_{i\mu\bar{\sigma}} c_{i\nu\sigma}, \nonumber
\end{eqnarray}
where $c_{i\mu\sigma}^\dagger (c_{i\mu\sigma})$ represents the 
creation (annihilation) operator of $\mu$-orbital with 
$\sigma$-spin at the $i$-th C$_{60}$ molecule and 
$n_{i\mu\sigma}$ is its number operator.
The lines represent
the orbital energy, 
the intermolecular transfer energy between different orbitals,
that between the same orbitals, 
the intra-orbital Coulomb energy,  
the inter-orbital Coulomb energy and 
the $S_zS_z$ part of the exchange interaction, 
and the $S_+S_-$ part of exchange interaction, respectively.
The intra-orbital and inter-orbital Coulomb energies 
are assumed to be the same. 
LUMO$_{\rm z}$'s are not included because they are not occupied on 
any C$_{60}$'s in the molecular alignment shown in Fig~\ref{fig:c60_jt}. 
It should be noted that a small transfer energy exists even between 
different orbitals related to the third term 
because of a small deviation from the perfect orthorhombic symmetry 
of the crystal. 

We suggest that the unpaired electron can be regarded 
to exist locally on each molecule 
since the Coulomb energy is sufficiently larger 
than the transfer energy. 
The energies $t_\ell$ and $U$ are estimated 
to be about 0.05 eV and 0.6 eV, respectively\cite{Saito91PRL,Suzuki95PRB}.
Experimentally, the Curie-Weiss like behavior is also 
observed in the magnetic susceptibility 
even under the application of  pressure\cite{Mizoguchi00PRL}. 
For a localized system, the Hamiltonian (\ref{eq:Hubbard}) 
can be transformed into 
an extended Heisenberg Hamiltonian. 
\begin{eqnarray} 
H   &=& -J_1\sum_{i}\vec{S}_i \cdot \vec{S}_{i+1}
      -J_2\sum_{i}\vec{S}_i \cdot \vec{S}_{i+2},     \label{eq:Hamiltonian}\\
J_1 &=& -\frac{4}{U}t_g^2+\frac{4t_\ell^2 J}{(U+\Delta)^2}, \label{eq:J1}\\
J_2 &=&  \frac{-4t_\ell^4}{U(U+\Delta)^2},
\end{eqnarray}
where $J_1$ and $J_2$ represent intrachain exchange 
interactions between nearest neighbors and second nearest neighbors, 
respectively.  
The magnitude is calculated from 
a second order perturbation  of the transfer energy
and the maximum terms of a fourth order perturbation.  
Details of the transformation are reported for 
another orbital ordering system, K$_2$CuF$_4$\cite{Kawamoto97JPSJ}. 
From this Hamiltonian, the Curie temperature $T_{\rm C}$ is derived 
with mean field theory as 
\begin{eqnarray}
T_{\rm C}^{\rm MF} &=& \frac{2}{3k}S(S+1)J(0) \\
          &=& \frac{1}{k}(J_1+J_2+2J_3 ) , \label{eq:TC}
\end{eqnarray}
where we assume the existence of interchain ferromagnetic couplings $J_3$, and 
$k$ represents the Boltzmann factor. 
$S$(=1/2) and $J(0)(\equiv 2J_1+2J_2+4J_3)$ 
indicate the value of a single spin and the 
$q$=0 component of the Fourier transformation of exchange interactions $J(q)$, 
respectively. 
It should be noted that the mean field theory generally overestimates the  
Curie temperature $T_{\rm C}$. 
Here we derive the pressure dependence of TDAE-C$_{60}$ 
based on Hamiltonian \ref{eq:Hamiltonian}. 
It is assumed that the transfer energy linearly depends 
on pressure as
$t_a=t_a^0+pt_a^\prime ( a = \ell, g). \label{eq:pressure}$
To describe the difference of the Curie temperature derived by 
mean field theory $T_{\rm C}^{\rm MF}$ 
from that observed experimentally $T_{\rm C}^{\rm Ex}$ , 
a reduction parameter $\alpha$ is introduced as 
$T_{\rm C}^{\rm Ex} = \alpha T_{\rm C}^{\rm MF}$. 

The solid line shown in Fig.~\ref{fig:p_tc} 
represents the simulation result of 
the pressure dependence of $T_{\rm C}$ with  
$t_\ell^0$=0.065eV, $t_g^0$=0.0035eV,
$t_\ell^\prime$=0.001eV/kbar, $t_g^\prime$=0.00243eV/kbar,
$U$=0.55eV, $J$=0.09eV, $\Delta$=0.15eV, $J_3$=3K, and $\alpha$=0.75. 
It is found that the theoretical result simulates the observation well. 
We emphasize that the parameters used in the simulation are reasonable. 
The intramolecular parameters $U$, $J$, and $\Delta$ are almost the same as the 
values estimated by Suzuki and Nakao\cite{Suzuki95PRB}.
The report of ab-initio calculations for fcc-C$_{60}$ by Saito and 
Oshiyama is used for the determination of 
the intermolecular transfer $t_\ell^0$\cite{Saito91PRL}.  
The width of the LUMO band is calculated to be about 0.5 eV in fcc-C$_{60}$ and 
the intermolecular transfer energy can be roughly estimated as 0.04eV. 
It is in good agreement with our parameter $t_\ell^0$. 
We have no quantitative information of the intermolecular interaction $J_3$. 
It is considered that 3K may be appropriate for the value of $J_3$  
because the interchain distance is also longer than the 
intrachain molecular distance by about 0.3~\AA. 
In conclusion, the parameters used in our simulation 
is found to be quite reasonable for TDAE-C$_{60}$. 
Tanaka {\it et al.} calculated the dependence of 
the intermolecular magnetic coupling between C$_{60}$'s on their orientation 
with a semi-empirical approach without the JT distortion\cite{Tanaka96CPL}. 
They reported that even the strongest ferromagnetic coupling is 
very small, about 0.03 K. 
Therefore, it is very important 
that the high $T_{\rm C}$ of this material can be explained 
by the JT distortion of C$_{60}$'s with the reasonable parameters. 

\begin{figure}
\begin{center}
  \leavevmode
  \epsfxsize=80mm
  \epsfbox{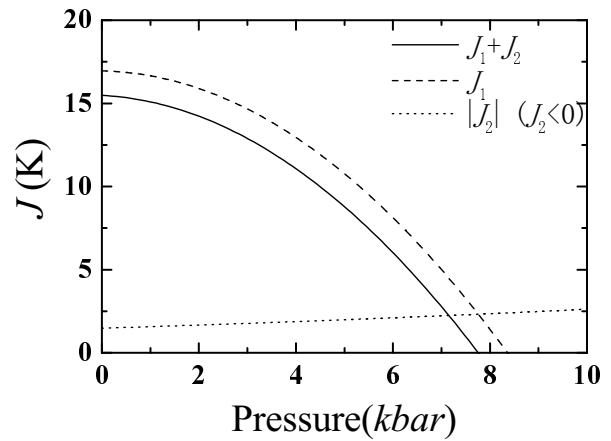}
\end{center}
\caption{
The pressure dependence of exchange interaction between intrachain 
nearest neighbors $J_1$, between second nearest neighbors $J_2$, 
and their sum. 
}
\label{fig:p_J}
\end{figure}

Figure~\ref{fig:p_J} shows the pressure dependence of the intermolecular 
exchange interactions $J_1$ and $J_2$ as well as their sum. 
It is found that the negative parabolic shape in the pressure dependence 
of $T_{\rm C}$ is almost determined by $J_1$ 
although its magnitude is weakened by the antiferromagnetic interaction $J_2$. 
The negative dependence of $T_{\rm C}$ on the application of pressure  
is formed by a coefficient of the parabolic term in the 
pressure dependence of $J_1$ shown in eq.\ref{eq:J1}. 
With the parameters used in the simulation, 
it is expected that $T_{\rm C}$ is decreased by applying pressure 
in the case of $t_\ell^\prime<12t_g^\prime$. 
The parameters for the simulation satisfy this condition. 
The crystal structure of TDAE-C$_{60}$ is slightly different 
from the orthorhombic one 
although the C$_{60}$ molecule has complete orthorhombic symmetry. 
Therefore, the enhancement of the crystal field at C$_{60}$'s site due to 
the application of pressure makes $t_g$ significantly larger. 
In fact, it is suggested from an ab initio calculation that 
$t_g^\prime$ and $t_\ell^\prime$ have the same order 
in another orbital ordering system, K$_2$CuF$_4$, 
in which local crystal field at the Cu site
also has a lower symmetry 
than that of JT distorted molecules\cite{Kawamoto97JPSJ}. 

A helical magnetism can be realized 
if the second neighbor coupling $J_2$ is sufficiently 
large\cite{Kawamoto97JPSJ}. 
We examine the stability of ferromagnetism against the helical magnetism. 
According to the mean field theory, the magnetic order with the wave number Q 
is realized if the Q-component is the largest in the Fourier form of an 
exchange interaction $J(q)$. 
In the case of $J_1/|J_2| > 4$, a ferromagnetic ordering is the most stable. 
Otherwise a helical magnetism is realized. 
It is evident in Fig.~\ref{fig:p_J} that helical ordering 
is expected above 6 kbar. 
It is not clarified  at the present time 
whether the magnetic ordering at $p$=7.4 kbar is 
ferromagnetic or helical ordering with a long wavelength. 

Let us discuss the dimensionality of TDAE-C$_{60}$ and 
the origin of the interchain ferromagnetic interaction. 
We assume the one-dimensionality of this system in our model. 
It is also indicated experimentally with 
the measurement of the electrical conductivity\cite{Omerzu97SynthM}. 
Although 
Blinc {\it et al.} reported that TDAE-C$_{60}$ is the isotropic 
ferromagnet in the temperature dependence of the spin-wave resonance in 
ESR\cite{Blinc98PRB}, 
the temperature dependence of the magnetization of
the quasi-one dimensional system with $J_1/J_3\sim6$ is not significantly 
different from that of an isotropic system. 
Therefore, the assumption of quasi-one-dimensionality 
does not conflict with the observations by Blinc {\it et al.}. 
The origin of interchain ferromagnetic coupling can be explained 
with the  orbital ordering structure shown in Fig.~\ref{fig:c60_jt}. 
In this structure, the ferromagnetic ordering is favored not only 
inside the chain but also interchain coupling\cite{Kawamoto97SSC}. 
The quasi-one dimensionality can be explained with this mechanism
since the magnitude of magnetic interaction depends on intermolecular
transfer energy. 
The interchain distance is about 0.3~\AA~longer than the intrachain molecular 
distance. 
Another model is the superexchange mechanism by way of TDAE molecules. 
The status of unpaired electrons on TDAE$^+$ cations is an open question 
but it is likely that their molecular spins form pairs\cite{Pokhodnia99JCP}. 
The reason for electron pairing on TDAE should be clarified 
in order to discuss the interchain magnetic interaction. 

In this paper, we assumed that one two-fold axis is parallel to the $c$-axis.
However, the constraint for realizing the ferromagnetic interaction along the 
$c$-axis is looser in reality. 
The suppression of the electronic transfer interaction between singly occupied 
molecular orbitals of neighboring C$_{60}$ anions is the most important. 
Consequently, our model requires that the elongated axes 
of neighboring C$_{60}$'s be oriented perpendicular 
or parallel to the c-axis and also perpendicular to each other.  
A structure proposed with X-ray analysis\cite{Narymbetov00Nature} 
is feasible with this constraint. 

Recently, additional collateral evidence has been reported. 
Kambe {\it et al.} also observed a structural phase transition at 180 K with 
X-ray diffraction measurements and  
they have reported the possibility of cooperative JT ordering 
in the low-temperature phase\cite{Kambe00PRB}. 
Clear experimental evidence of D$_{\rm 2h}$ JT distortion 
in monoanion-C$_{60}$ was also reported 
on a single crystal of a model compound of 
monoanion-C$_{60}$, [As(C$_6$H$_5$)$_4$]$_2$C$_{60}$Cl\cite{Bietsch00CPL}. 
Furthermore, in another C$_{60}^-$ compound, 
an X-ray analysis also suggests the stabilization of the static
JT distortion by symmetry lowering of the crystal caused 
by rotational ordering\cite{Launois00EPJB}. 
For complete confirmation of our model, 
we urge to perform the neutron scattering observation. 
In neutron scattering measurements, we can obtain information on the 
spin density. 
The spin distribution on C$_{60}^-$'s is not spherical but 
takes large values along the belt around the elongated axis, 
as shown in Fig.~\ref{fig:c60_jt}. 
It has a much larger spatial difference than the lattice distortion. 

The authors gratefully thank Professor N. Suzuki and Dr. S. Suzuki 
for valuable comments. 
We also acknowledge Professor D. Mihailovic, Dr. B. Narymbetov 
and Dr. A. Omerzu for useful discussions. 
This work was partly supported by a Grant-in-Aid for
Scientific Research on the Priority Area `Fullerenes
and Nanotubes' by the Ministry of 
Education, Sports, Culture, Science and Technology of Japan.

\end{document}